\documentclass[twoside,twocolumn,aps,pra,floatfix]{revtex4}
\usepackage[hypertex]{hyperref}
\usepackage{amsmath}
\usepackage{graphicx}
\usepackage{bm}
\usepackage{epsfig}
\usepackage{amsfonts}
\usepackage{amssymb}
\usepackage{mathrsfs}

\def\wgt{\mathop{\rm wgt}}
\def\Hn{\mathcal{H}_2^{\otimes n}}
\def\H#1{\mathcal{H}_2^{\otimes #1}}
\def\clg{\mathop{\mathrm{Cl}_{\mathcal{G}}}}

\def\Z{{\overline Z}}

\def\X{{\overline X}}
\newcommand{\bra}[1]{\left\langle{#1}\right\vert}
\newcommand{\ket}[1]{\left\vert{#1}\right\rangle}
\newcommand{\ts}{\textstyle}

\def\bplus{\mathbin{\boxplus}}
\def\bigbplus{\mathop{\raisebox{-0.75ex}{\text{\LARGE{$\boxplus$}}}}\limits}

\newtheorem{theorem}{Theorem}

\begin{document}
\advance\belowcaptionskip by -0.05in
\setcounter{dbltopnumber}{3}

\title{Structured Error Recovery for Codeword-Stabilized Quantum Codes}
\author{Yunfan~Li}
\email{yunfan@ee.ucr.edu}
\affiliation{Department of Electrical Engineering, University of California,
Riverside, CA, 92521, USA}
\author{Ilya~Dumer}
\email{dumer@ee.ucr.edu}
\affiliation{Department of Electrical Engineering, University of California,
Riverside, CA, 92521, USA}
\author{Markus~Grassl}
\email{markus.grassl@nus.edu.sg}
\affiliation{Centre for Quantum Technologies, National University of
  Singapore, Singapore 117543, SINGAPORE}
\author{Leonid~P.~Pryadko}
\email{leonid@landau.ucr.edu}
\affiliation{Department of Physics \& Astronomy,
University of California, Riverside, CA, 92521, USA}

\date{\today}
\begin{abstract}
  Codeword stabilized (CWS) codes are, in general, non-additive quantum
  codes that can correct errors by an exhaustive search of
  different error patterns, similar to the way that we decode
  classical non-linear codes.  For an $n$-qubit quantum code correcting
  errors on up to $t$ qubits, this brute-force approach consecutively
  tests different errors of weight $t$ or less, and employs a separate
  $n$-qubit measurement in each test.  In this paper, we suggest an
  error grouping technique that allows 
  to simultaneously test large groups of
  errors in a single measurement.  This structured error
  recovery technique exponentially reduces the number of measurements
  by about $3^t$ times.  While it still leaves exponentially many
  measurements for a generic CWS code, the technique is equivalent to
  syndrome-based recovery for the special case of additive CWS codes.
\end{abstract}

\maketitle

\section{Introduction}

Quantum computation makes it possible to achieve polynomial complexity
for many classical problems that are believed to be hard
\cite{Shor-factoring-1994,Nielsen-book}. To preserve coherence,
quantum operations need to be protected by quantum error correcting
codes (QECCs)
\cite{shor-error-correct,Knill-Laflamme-1997,Bennett-1996}.  With
error probabilities in elementary gates below a certain threshold, one
can use multiple layers of encoding (concatenation) to reduce errors
at each level and ultimately make arbitrarily-long quantum computation
possible \cite{Knill-error-bound,Rahn-2002,Steane-2003,%
  Fowler-QEC-2004,Fowler-2005,fowler-thesis-2005,%
  Knill-nature-2005,Knill-2005,Raussendorf-Harrington-2007}.

The actual value of the threshold error probability strongly depends
on the assumptions of the error model and on the chosen architecture,
and presently varies from $10^{-3}\%$ for a chain of qubits with
nearest-neighbor couplings \cite{Stephens-Evans-2009} and $
0.7\%$ for qubits with nearest-neighbor couplings in two
dimensions \cite{Raussendorf-Harrington-2007}, to $3\%$ with
postselection \cite{Knill-nature-2005}, or even above $10\%$ if
additional constraints on errors are imposed \cite{Knill-2005}.

The quoted estimates have been made using  stabilizer codes,
an important class of codes which originate from additive quaternary
codes, and have a particularly simple structure based on Abelian
groups \cite{gottesman-thesis,Calderbank-1997}.  Recently, a more
general class of codeword stabilized (CWS) quantum codes was
introduced in
Refs.~\cite{Smolin-2007,Chen-Zeng-Chuang-2008,Cross-CWS-2009,Chuang-CWS-2009}.
This class includes stabilizer codes, but is more directly related to
non-linear classical codes.

This direct relation to classical codes is, arguably, the most
important advantage of the CWS framework.  Specifically, the classical
code associated with a given CWS quantum code has to correct certain
error patterns induced by a graph associated with the code.  The graph
also determines the graph state \cite{Hein-2006} serving as a starting
point for an encoding algorithm exploiting the structure of the
classical code \cite{Cross-CWS-2009}.  With the help of powerful
techniques from the theory of classical codes, already several new
families of non-additive codes have been discovered, including codes
with parameters proven to be superior to any stabilizer code
\cite{Smolin-2007,Chen-Zeng-Chuang-2008,Cross-CWS-2009,%
  Yu-Chen-Lai-Oh-2008,%
  Grassl-Roetteler-2008A,Grassl-Roetteler-2008B,Grassl-2009}.

Both classical additive codes and additive quantum codes can be
corrected by first finding the syndrome of a corrupted vector or
quantum state, respectively, and then looking up the corresponding
error (coset leader) in a precomputed table \cite{MS-book}.  This is
not the case for non-linear codes.  In fact, even the notions of a
syndrome and a coset become invalid for general non-linear codes.
Furthermore, since quantum error correction must preserve the original
quantum state in all intermediate measurements, it is more restrictive
than many classical algorithms.  Therefore, the design of a useful CWS
code must be complemented by an efficient quantum error correction
algorithm.

The goal of this work is to address this important unresolved problem
for binary CWS codes.  First, we design a procedure to {\em detect\/}
an error in a narrower class, the \textit{union stabilizer} (USt)
codes, which possess some partial group structure
\cite{Grassl-1997,Grassl-Roetteler-2008A,Grassl-Roetteler-2008B}.
Then, for a general CWS code and a set of graph-induced maps of
correctable errors forming a group, we construct an auxiliary USt code
which is the union of the images of the original CWS code shifted by
all the elements of the group.  Finally, we construct Abelian groups
associated with correctable errors located on certain {\em index
  sets\/} of qubits.  The actual error is found by first applying
error-detection to locate the index set with the relevant auxiliary
USt code, then using a collection of smaller USt codes to pinpoint the
error in the group.  Since we process large groups of errors
simultaneously, we make a significant reduction of the number of
measurements compared with the brute force error correction for
non-linear (quantum or classical) codes.

More precisely, we consider an arbitrary distance-$d$ CWS code
$((n,K,d))$ that uses $n$ qubits to encode a Hilbert space of
dimension $K$ and can correct all $t$-qubit errors, where $t=\lfloor
(d-1)/2\rfloor$.  In Sec.~\ref{sec:background} we give a brief
overview of the notations and relevant facts from the theory of
quantum error correction.  Then in Sec.~\ref{sec:generic}, we
construct a reference recovery algorithm that deals with errors
individually.  This algorithm requires up to $B(n,t)$ measurements,
where
\begin{equation}
  B(n,t)\equiv \sum_{i=0}^{t}\ts \binom{n}{i}\,3^{i}
  \label{eq:sphere}
\end{equation}
is the total number of errors of size up to $t$ (this bound is tight
for non-degenerate codes).  Each of these measurements requires up to
$n^2+K\mathcal{O}(n)$ two-qubit gates.  In order to eventually reduce
the overall complexity, we consider the special case of USt codes in
Sec.~\ref{sec:ust-measurement}.  Here we design an error-detecting
measurement for a USt code with a translation set of size $K$ that
requires $\mathcal{O}(K n^2)$ two-qubit gates to identify a single
error.  Our error grouping technique presented in
Sec.~\ref{sec:clusters} utilizes such a measurement to check for
several errors at once.  For additive CWS codes the technique reduces
to stabilizer-based recovery [Sec.~\ref{sec:additive}].  In the case
of generic CWS codes [Sec.~\ref{sec:generic-cws}], we can
simultaneously check for all errors located on size-$t$ qubit
clusters; graph-induced maps of these errors form groups of size up to
$2^{2t}$.  Searching for errors in blocks of this size requires up to
$\binom{n}{t}-1$ measurements to locate the cluster, plus up to $2t$
additional measurements to locate the error inside the group.  In
Sec.~\ref{sec:conclusions} we discuss the obtained results and outline
the directions of further study.  Finally, in
Appendix~\ref{app:orthogonality} we consider some details of the
structure of corrupted spaces for the codes discussed in this work.

Note that some of the reported results have been previously announced
in Ref.~\cite{Li-Dumer-Pryadko-2009}.

\section{Background}\label{sec:background}
\subsection{Notations}
\begin{figure}[ptbh]
\centering
\includegraphics[scale=1.0]{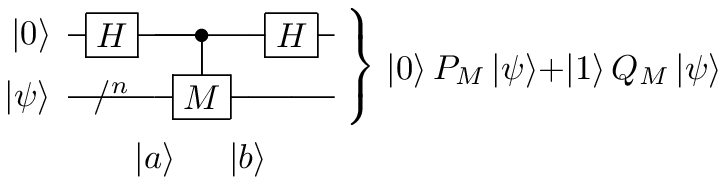}
\caption{Measurement of an observable $M$ with all eigenvalues $\pm1$.
  The first Hadamard gate prepares the ancilla in the state
  $(\ket0+\ket1)\ket\psi/\sqrt2$, hence
  $\ket{a}=(\ket0+\ket1)\ket\psi/\sqrt2$.  The controlled-$M$ gate
  returns $\ket{b}=C^M\ket{a}= (\ket0\ket\psi+\ket1
  M\ket\psi)/\sqrt2$.  The second Hadamard gate finishes the
  incomplete measurement,
  $\ket{c}=\ket0{P}_M\ket\psi+\ket1{Q}_M\ket\psi$, where we used the
  projector identities~(\ref{eq:projector-identities}).  If the
  outcome of the ancilla measurement is $\ket{0}$, the result is the
  projection of the initial $n$-qubit state $|\psi\rangle$ onto the
  $+1$ eigenspace of $M$ (${P}_M\ket\psi$), otherwise it is the
  projection onto the $-1$ eigenspace of $M$ (${Q}_M\ket\psi$).  For
  an input state $\ket 1\ket\psi$ with ancilla in the state $\ket1$,
  the circuit returns $\ket1{P}_M\ket\psi +\ket0{Q}_M\ket\psi$.}
\label{fig:measurementM}
\end{figure}

Throughout the paper, $\mathcal{H}_{2}=\mathbb{C}^2$ denotes the
complex Hilbert space that consists of all possible states
$\alpha|0\rangle+\beta|1\rangle$ of a single qubit, where
$\alpha,\beta\in\mathbb{C}$ and $\left\vert \alpha\right\vert
^{2}+\left\vert \beta\right\vert ^{2}=1.$ Correspondingly, we use the
space $\Hn=(\mathbb{C}^2)^{\otimes n}=\mathbb{C}^{2^n}$ to represent
any $n$-qubit state. Also,
\begin{equation}
  \mathscr{P}_{n}\equiv i^m\{I,X,Y,Z\}^{\otimes n},\quad\text{$m=0,\ldots,3$}
  \label{eq:Pauli-group}
\end{equation}
denotes the Pauli group of size $2^{2n+2}$, where $X$, $Y$, $Z$ are
the usual (Hermitian) Pauli matrices and $I$ is the identity matrix.
The members of this group are called Pauli operators; the operators in
(\ref{eq:Pauli-group}) with $m=0$ form a basis of the vector space
that consists of all operators acting on $n$-qubit states.  The {\em
  weight\/} $\wgt(E)$ of a Pauli operator $E$ is the number of terms
in the tensor product~(\ref{eq:Pauli-group}) which are not a scalar
multiple of identity.  Up to an overall phase, a Pauli operator can be
specified in terms of two binary strings, ${\bf v}$ and ${\bf u}$,
$$
U= Z^{\bf v}X^{\bf u}\equiv Z_1^{v_1} Z_2 ^{v_2}\ldots Z_n ^{v_n} X_1^{u_1} X_2 ^{u_2}\ldots X_n ^{u_n}.
$$
Hermitian
operators in $\mathscr{P}_{n}$ have eigenvalues equal to $1$ or $-1$.
Generally, unitary operators (which can be outside of the Pauli group) which
are also Hermitian, i.e., all eigenvalues are $\pm1$, will be
particularly important in the discussion of measurements.  We will
call these operators {\em measurement operators}.  Indeed, for such an
operator $M$, a measurement gives a Boolean outcome and can be
constructed with the help of a single ancilla, two Hadamard gates, and
a controlled $M$ gate \cite{gottesman-thesis} (see
Fig.~\ref{fig:measurementM}).  The algebra of measurement operators is
related to the algebra of projection operators discussed in
 \cite{Aggarwal-Calderbank-2008}, but the former operators, being
unitary, are more convenient in circuits.

A measurement of an observable defined by a Pauli operator $M$ will be
also called Pauli measurement \cite{Elliott-2008}.  For lack of a
better term, other measurements will be called \emph{non-Pauli};
typically the corresponding circuits are much more complicated than
those for Pauli measurements.

We say that a state $|\psi\rangle\in{\Hn}$ is stabilized
(anti-stabilized) by a measurement operator $M$ if
$M|\psi\rangle=|\psi\rangle$ ($M|\psi\rangle=-|\psi\rangle)$.  The
corresponding projectors onto the positive and negative eigenspace are
denoted by $P_M$ and $Q_M$, respectively; they satisfy the identities
\begin{equation}
M=P_M-Q_M=2P_M-\openone=\openone-2Q_M.
\label{eq:projector-identities}
\end{equation}

We say that a space $\mathcal{Q}$ is stabilized by a set of operators
$\mathcal{M}$ if each vector in $\mathcal{Q}$ is stabilized by each
operator in $\mathcal{M}$.  We use $\mathcal{P}(\mathcal{M})$ to
denote the maximum space stabilized by $\mathcal{M}$, and
$\mathcal{P}^\perp(\mathcal{M})$ to denote the corresponding
orthogonal complement.  For a set $\mathcal{M}$ of measurement
operators, each state in $\mathcal{P}^\perp(\mathcal{M})$ is
anti-stabilized by some operator in $\mathcal{M}$.

When discussing complexity, we will quote the two-qubit complexity
which just counts the total number of two-qubit gates.  Thus, we
ignore any communication overhead, as well as any overhead associated
with single-qubit gates.  For example, the complexity of the
measurement in Fig.~\ref{fig:measurementM} is just that of the
controlled-$M$ gate operating on $n+1$ qubits \cite{Barenco-1995}.  For
all circuits we discuss, the total number of gates (single- and
two-qubit) is of the same order in $n$ as the two-qubit complexity.

\subsection{General QECCs}\label{sec:general}
A general $n$-qubit quantum code $\mathcal{Q}$ encoding $K$ quantum
states is a $K$-dimensional subspace of the Hilbert space $\Hn$.  Let
$\{|i\rangle\}_{i=1}^{K}$ be an orthonormal basis of the
$K$-dimensional code $\mathcal{Q}$ and let
$\mathcal{E}\subset\mathscr{P}_n$ be some set of
Pauli errors.  The overall phase of an error [$i^m$ in
  Eq.~(\ref{eq:Pauli-group})] is irrelevant and will be largely
ignored.  The code detects all errors $E\in\mathcal{E}$ if and only if
\cite{Nielsen-book,gottesman-thesis}
\begin{equation}
\langle j|E|i\rangle=C_{E}\delta_{ij} \label{SNcondtion-detecting}
\end{equation}
where $C_{E}$ only depends on the error $E$, but is independent of the
basis vectors \cite{endnote38}.  The code has distance $d$ if it can
detect all Pauli errors of weight $(d-1)$, but not all errors of weight
$d$.  Such a code is denoted by $((n,K,d))$.

The necessary and sufficient condition for correcting errors in
$\mathcal{E}$ is that all non-trivial combinations of errors from
$\mathcal{E}$ are detectable.  This
gives \cite{Knill-Laflamme-1997,Bennett-1996}
\begin{equation}
\langle j|E_{1}^{\dag}E_{2}|i\rangle=C_{E_{1}, E_{2}}\delta_{ij},
\label{SNcondtion-correcting}
\end{equation}
where $E_{1},E_{2}\in\mathcal{E}$ and, again, $C_{E_{1}, E_{2}}$ is
the same for all basis states $i$, $j$.  A distance-$d$ code corrects
all errors of weight $s$ such that $2s\le d-1$, that is, $s\le t\equiv \lfloor
(d-1)/2\rfloor$.

The code is \emph{non-degenerate} if linearly independent errors from
${\cal E}$ produce corrupted spaces $E({\cal Q})\equiv
\{E\ket\psi:\ket\psi\in{\cal Q}\}$ whose intersection is trivial
(equals to $\{0\}$); otherwise
the code is {\em degenerate} \cite{Calderbank-1997}.  A stricter
condition that the code is \emph{pure} (with respect to $\mathcal{E}$)
requires that the corrupted spaces $E_1({\cal Q})$ and $E_2({\cal Q})$
be mutually orthogonal for all linearly independent correctable
errors $E_1$, $E_2\in{\cal E}$.

For a degenerate code, we call a pair of correctable errors
$E_1,E_2\in\mathcal{E}$ {\em mutually-degenerate\/} if the corrupted
spaces $E_1({\cal Q})$ and $E_2({\cal Q})$ coincide.
Such errors belong to the same {\em degeneracy class\/}.
For recovery, one only
needs to identify the degeneracy class of the error that happened.
The operators like $E_1^\dagger E_2$, connecting mutually-degenerate
correctable errors $E_1$ and $E_2$, have no effect on the code and can
be ignored.

As shown in Appendix~\ref{app:orthogonality}, for all codes discussed
in this work, any two correctable errors $E_1$, $E_2$ yield 
corrupted spaces $E_1({\cal Q})$, $E_2({\cal Q})$ that are either
identical or orthogonal.  Then, errors from different degeneracy
classes take the code to corrupted spaces that are mutually
orthogonal.  Also, for these codes, a non-degenerate code is always
pure.  In terms of the error correction
condition~(\ref{SNcondtion-correcting}), we have $C_{E_1,E_2}=0$ for
errors $E_1$, $E_2$ in different degeneracy classes and
$C_{E_1,E_2}\neq 0$ for errors in the same degeneracy class.

\subsection{Stabilizer codes}\label{sec:stabilizer-codes}
Stabilizer codes \cite{gottesman-thesis} are a well known family of
quantum error-correcting codes that are analogous to classical linear
codes. An $[[n,k,d]]$ stabilizer code maps a $2^{k}$-dimensional
$k$-qubit state space into a $2^{k}$-dimensional subspace of an
$n$-qubit state space.

The code is defined as the space stabilized by an Abelian subgroup of the
$n$-qubit Pauli group, $\mathscr{S}\subset\mathscr{P}_n$, with $n-k$ Hermitian
generators, $\mathscr{S}=\langle G_1,\ldots, G_{n-k}\rangle$.  For such a space
to exist, it is necessary that $-\openone\notin\mathscr{S}$.  The Abelian group
$\mathscr{S}$ is called the \textit{stabilizer} of $\mathcal{Q}$.  Explicitly,
\begin{equation}
  \mathcal{Q}\equiv \left\{|\psi\rangle:S|\psi\rangle=|\psi\rangle,\;
    \forall\,
    S\in\mathscr{S}\right\}.
\label{eq:stabilizer-code-defined}
\end{equation}
The code $\mathcal{Q}$ is stabilized by $\mathscr{S}$ iff it is
stabilized by all $n-k$ generators $G_{i}$.  In other words, it  is an
intersection of subspaces stabilized by $G_i$,
\begin{equation}
  \label{eq:stabilizer-code-intersection}
  \mathcal{Q}=\bigcap_{i=1}^{n-k}\mathcal{P}(G_i).
\end{equation}

The {\em normalizer\/} of $\mathscr{S}$ in $\mathscr{P}_n$, denoted as
$\mathscr{N}$, is the group of all Pauli operators $U$ which fix
$\mathscr{S}$ under conjugation ($U^\dagger S U=S$ for all
$S\in\mathscr{S}$).  The term {\em normalizer\/} reflects the fact
that these operators commute with $\mathscr{S}$
\cite{gottesman-thesis}.  It is possible to construct $2k$ logical
operators ${\overline X}_j$, ${\overline Z}_j$, $j=1,\ldots,k$, with
the usual commutation relations, that together with the generators of
$\mathscr{S}$ generate the normalizer (modulo an overall phase factor)
\cite{gottesman-thesis,Wilde-2009}.
The $(n-k)$ generators $G_i$ of the stabilizer, along with the $k$
operators ${\overline Z}_j$, generate a subgroup $\mathscr{S}\equiv
\langle G_1,\ldots,G_{n-k},{\overline Z}_1,\ldots {\overline
 Z}_k\rangle$ of $\mathscr{P}_n$ which becomes a maximal Abelian
subgroup when including the generator $i\openone$.  The group
$\mathscr{S}$ stabilizes a unique state
\begin{equation}
|s\rangle\equiv
|\overline{0\ldots 0}\rangle.\label{eq:stabilized-state}
\end{equation}
The operators
${\overline X}_j$ acting on $|s\rangle$ generate the basis of the
code,
\begin{equation}
  |\overline{c_1\ldots c_k}\rangle \equiv {\overline
  X}_1^{c_1}\ldots {\overline X}_k^{c_k}|s\rangle.\label{eq:stabilizer-code-basis}
\end{equation}

Generally, each detectable Pauli error $E_{j}\in\mathscr{P}_n$ that
acts non-trivially on the code anti-commutes with at least one
generator $G_{i}$, and errors from different degeneracy classes
anti-commute with different subsets of $\mathscr{S}$.  We can thus
identify a degeneracy class by the set of generators $G_i$ which
anti-commute with it.  The corrupted code space
$E_{j}(\mathcal{Q})=\{E_{j} |\psi\rangle:|\psi\rangle\in\mathcal{Q}\}$
is anti-stabilized by those generators $G_{i}$ that anti-commute with
$E_{j}$. Indeed,
\[
G_{i}(E_{j}|\psi\rangle)=(-E_{j}G_{i})|\psi\rangle=-E_{j}|\psi\rangle,
\]
which means that the measurement $G_{i}$ of gives $-1$.  By measuring all
$G_{i}$, we get the \emph{syndrome} that consists of $n-k$ numbers $1$ or
$-1$.  There are in total $ 2^{n-k}$ possible syndromes identifying
different error degeneracy
classes, including the trivial error $\openone$ which corresponds to the
all-one syndrome vector.

Any two corrupted spaces $E_{i}(\mathcal{Q})$ and $E_{j}(\mathcal{Q})$
are mutually orthogonal or identical.  The whole
$2^{n}$-dimensional $n$-qubit state space $\Hn$ is thus divided into
$L\equiv 2^{n-k}$ orthogonal $2^{k}$-dimensional subspaces
${\mathcal{Q}_{j}\equiv}E_{j}(\mathcal{Q})$,
\begin{equation}
  {\Hn}=\bigoplus_{j=0}^{L-1}{\mathcal{Q}_{j}}, \quad
  {\mathcal{Q}_{i}}\perp
 \mathcal{Q}_j\;\,\text{for}\;\,
  i\neq j.
  \label{eq:subspace-decomposition}
\end{equation}
The representatives of different error classes can be chosen to
commute with each other and with the logical operations ${\overline
  X}_i$.  These representatives form an Abelian group \cite{endnote1}
$\mathscr{T}\equiv\langle
g_1,\ldots g_{n-k}\rangle$ whose generators $g_i$ can be chosen to
anti-commute with only one of the generators of the stabilizer each,
$g_i G_j=(-1)^{\delta_{ij}} G_j g_i$ (this follows from Proposition
10.4 in Ref.~\cite{Nielsen-book}).  Altogether, the
generators
$\{g_1,\ldots, g_{n-k}\}$ can be regarded as a
set of Pauli operators forming the basis of the cosets of the
normalizer $\mathscr{N}$ of the code $\mathcal{Q}$ in $\mathscr{P}_n$.

\textbf{Example 1.}
The $[[5,1,3]]$ stabilizer code is defined by the
generators
\begin{equation}
\begin{split}
G_{1}=& XZZXI,\quad G_{2}=IXZZX, \\
G_{3}=& XIXZZ,\quad G_{4}=ZXIXZ.
\end{split}\label{eq:stab513}
\end{equation}
For this code, the logical operators can be taken as
\begin{equation}
{\overline{X}}=ZZZZZ,\quad {\overline{Z}}=XXXXX.  \label{eq:logical513}
\end{equation}
A basis of the code space is (up to normalization)
\begin{equation*}
|\bar{0}\rangle =\prod_{i=1}^{4}{(\openone+G_{i})}|00000\rangle ,\quad
|\bar{1}\rangle ={\overline{X}}|\bar{0}\rangle.
\end{equation*}
By construction, both basis states are stabilized by the generators
$G_{i}$.  The corresponding stabilizer group is $\mathscr{S} =\langle
G_{1},\ldots ,G_{4}\rangle $.  The group of equivalence classes of
correctable errors is generated by the representatives (note the mixed
notation, e.g., $Z_1Z_3\equiv ZIZII$)
\begin{equation}
  g_1=Z_1Z_3,\; g_2=ZZZZI,\; g_3=ZZIZZ,\; g_4=Z_2Z_5.
  \label{eq:513-errors}
\end{equation}
The $g_j$ are chosen to commute with the logical operators and also to
satisfy $G_i g_j=(-1)^{\delta_{ij}}g_j G_i$.  Note that the operators
of weight one forming the correctable error set do not by themselves
form a group.  The generators $g_i$ can be used to map correctable
errors to the corresponding group elements with the same syndrome.
This gives, e.g., $Z_2\to g_2 g_4=Z_2\overline X$, $X_2\to
g_1=Z_{1}Z_{3}$, $Y_2\to g_1g_2g_4=Z_{1}Z_2 Z_{3}\overline X=Z_4Z_5$.
\hfill $\Box $

\subsection{Union Stabilizer codes}\label{sec:ust}
The decomposition~(\ref{eq:subspace-decomposition}) can be viewed as a
constructive definition of the Abelian group $\mathscr{T}\equiv
\langle g_1,\ldots g_{n-k}\rangle$ of all {\em translations\/} of the
original stabilizer code $\mathcal{Q}$ in $\mathscr{P}_n$.  (In the
following, this stabilizer code is denoted ${\cal Q}_0$.)  
Any two non-equivalent translations $t_i,t_j\in\mathscr{T}$ belong to
different cosets of the normalizer $\mathscr{N}_0$ of the code
$\mathcal{Q}_0$ in $\mathscr{P}_n$ and, therefore, yield mutually
orthogonal
shifts%
\begin{equation}
t_i(\mathcal{Q}_0)\perp t_j(\mathcal{Q}_0),\quad i\neq
j.\label{eq:ust-orthogonality} 
\end{equation}
A union stabilizer code
\cite{Grassl-1997,Grassl-Roetteler-2008A,Grassl-Roetteler-2008B}
(USt) is a direct sum%
\begin{equation}
  \label{eq:ust-defined}
  \mathcal{Q}\equiv \bigoplus_{i=1}^K t_i (\mathcal{Q}_0)
\end{equation}
of $K$  shifts of the code $\mathcal{Q}_0$ by 
non-equivalent translations $t_i\in\mathscr{T}$, $i=1,\ldots K$.

The basis of the code defined by (\ref{eq:ust-defined}) is the
union of the sets of basis vectors of all $t_i(\mathcal{Q}_0)$. As
a result, the dimension of $\mathcal{Q}$ is $K \,2^k$.  This code
is then denoted $((n,K 2^k, d))$, where $d$ is the distance of the
new code.  Generally, this distance does not exceed the distance
of the original code with respect to the same error set,
$d(\mathcal{Q})\le d(\mathcal{Q}_0)$.  However, if the code
$\mathcal{Q}$ is degenerate and the original code $\mathcal{Q}_0$
is one-dimensional, this need not be true.

\subsection{Graph states}
The unique state $|s\rangle$ defined by
Eq.~(\ref{eq:stabilized-state}) itself forms a single-state stabilizer
code $[[n,0,d']]$.  Its stabilizer $\mathscr{S}\equiv \langle
G_1,\ldots,G_{n-k},{\overline Z}_1,\ldots {\overline Z}_k\rangle $ has
exactly $n$ mutually commuting generators.  Note that for stabilizer
states we follow the convention that the distance $d'$ is
given by the minimum weight of the non-trivial elements of the
stabilizer $\mathscr{S}$.  Generally, such a state stabilized by an
Abelian subgroup in $\mathscr{P}_n$ of order $2^n$ (which does not
include $-\openone$) is called a {\em stabilizer state}
\cite{gottesman-thesis}.

A {\em graph state\/} \cite{Hein-2006} is a stabilizer state of a
group whose $n$ mutually commuting generators $S_i$ are defined by a
(simple) $n$-vertex graph $\mathcal{G}$ with an adjacency matrix
$R\in\{{0,1\}}^{n\times n}$.  Specifically, the generators are
\begin{equation}
  S_i\equiv X_i Z_1^{R_{i1}} Z_2^{R_{i2}}\ldots  Z_n^{R_{in}}\equiv
X_i Z^{{\bf r}_{i}},
\label{eq:canonical-generators}
\end{equation}
where ${\bf r}_i$, $i=1,\ldots n$ denotes the $i$th row of the
adjacency matrix $R$.  A graph state can be efficiently generated
\cite{Raussendorf-2003,Cross-CWS-2009} by first initializing every
qubit in the state $\ket{+}=(\ket{0}+\ket{1})/\sqrt{2}$ (e.g.,
applying the Hadamard gate on $|0\rangle$), and then using a
controlled-$Z$ gate $C_{i,j}^Z$ on every pair $(i,j)$ of qubits
corresponding to an edge of the graph $\mathcal{G}\equiv (V,E)$,
\begin{equation}
  |s\rangle=\prod_{(i,j)\in E}{C_{i,j}^{Z}H^{\otimes
      n}|0\rangle}_{n}\equiv
  U_{\mathcal{G}}|0\rangle_{n}, \label{stabilizedstate}
\end{equation}
where $|0\rangle_{n}$ is a state with all $n$ qubits in state
$|0\rangle$.

Any stabilizer state is locally Clifford-equivalent (LC-equivalent) to
a graph state
 \cite{Grassl-Klappenecker-Roetteler-2002,Schlingemann-2002,VandenNest-2004}.
That is, any stabilizer state can be transformed into a graph state by
individual discrete rotations of the qubits. 
This defines the
canonical form of a stabilizer state.

\begin{figure}[htb]
  \includegraphics[scale=1]{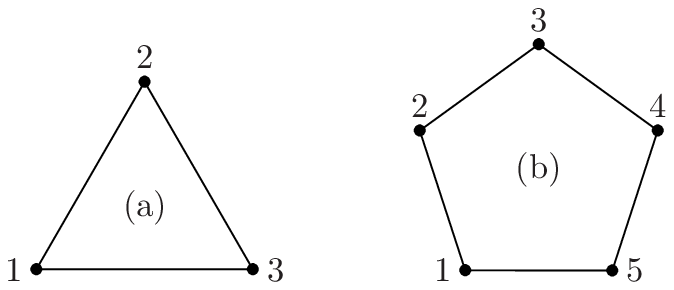}
  \caption{Ring graphs
    with (a) $3$ vertices and (b) $5$ vertices.}
\label{fig:tria}
\end{figure}

\textbf{Example 2.}
Consider the ring graph for $n=3$ [Fig.~\ref{fig:tria}(a)] which
defines the stabilizer generators $S_1=XZZ$, $S_2=ZXZ$, $S_3=ZZX$.
The corresponding stabilizer state $\ket s$ is an equal superposition
of all $2^3$ states [result of the Hadamard gates in
  Eq.~(\ref{stabilizedstate})], taken with positive or negative signs
depending on the number of pairs of ones at positions connected by the
edges of the graph [result of the gates $C_{i,j}^Z=(-1)^{ij}$].  In
the following expressions we omitted normalization for clarity:
\begin{alignat}{5}
    &|s\rangle  &{}={}&
      |000\rangle+|001\rangle+|010\rangle-|011\rangle\nonumber\\
      &&{}+{}& |100\rangle-|101\rangle-|110\rangle-|111\rangle
      \label{eq:s3}\\
   = S_2&|s\rangle  &{}={}&
      |010\rangle-|011\rangle+|000\rangle+|001\rangle\nonumber\\
      &&{}-{}&|110\rangle-|111\rangle+|100\rangle-|101\rangle.
\end{alignat}

\subsection{CWS codes and their standard form}
\label{sec:cws-defined}
Codeword stabilized (CWS) codes
\cite{Chen-Zeng-Chuang-2008,Cross-CWS-2009} represent a general class
of non-additive quantum codes that also includes stabilizer codes.
They can be viewed as USt codes originating from a stabilizer state.
Any CWS code is locally Clifford-equivalent to a CWS code which
originates from a graph state \cite{Cross-CWS-2009}.

Specifically, a CWS code $((n,K,d))$ in {\em standard form\/}
\cite{Cross-CWS-2009} is defined by a graph $\mathcal{G}$ with $n$
vertices and a classical code $\mathcal{C}$ containing $K$ binary
codewords ${\bf c}_i$.  The originating stabilizer state is the graph
state $|s\rangle$ defined by $\mathcal{G}$, and the {\em codeword
  operators\/} [translations in Eq.~(\ref{eq:ust-defined})] have the
form $W_i\equiv Z^{{\bf c}_i}$, $i=1,\ldots K$.  For a CWS code in
standard form we use the notation
$\mathcal{Q}=(\mathcal{G},\mathcal{C})$.

An important observation made in Ref.~\cite{Cross-CWS-2009} is that a
single qubit error $X$, $Z$, or $Y$ acting on a code state
\begin{equation}
|w_i\rangle\equiv W_i|s\rangle\label{eq:codeword-def}
\end{equation}
is equivalent (up to a phase) to an error composed only of $Z$
operators.  This establishes the following mapping between multi-qubit
errors and classical binary errors.  First, let
$E_{i}=Z_{i}^{\{0,1\}}X_{i}$ be an error acting on the $i$th qubit of
$W_j|s\rangle$.  Then we see that
\[
E_{i}W_{j}|s\rangle=E_{i}W_{j}S_{i}|s\rangle=\pm(E_{i}S_{i})W_{j}|s\rangle,
\]
where the term $E_{i}S_{i}=Z_{i}^{\{0,1\}}Z^{\mathbf{r}_{i}}$ consists
only of operators $Z$.  The general mapping of an error $E=i^m
Z^{\mathbf{v}}X^{\mathbf{u}}$ from the error set
$\mathcal{E}\subset\mathscr{P}_{n}$ to a classical error vector in
$\{0,1\}^{n}$ is defined as
\begin{equation}
\clg(E)\equiv \mathbf{v}\oplus\bigoplus_{l=1}^{n}u_{l}{\mathbf{r}_{l}},
\label{binary}
\end{equation}
where $u_{l}$ is the $l$\,th component of the vector $\mathbf{u}$.  We
will refer to both the binary vector $\clg(E)$ and the operator
$Z^{\clg(E)}$ as the {\em graph image\/} of the Pauli operator $E$.

Theorem 3 from Ref.~\cite{Cross-CWS-2009} establishes the correspondence between the error-correcting properties of a
quantum code $\mathcal{Q}$ and those of the corresponding classical
code $\mathcal{C}$.  It states that a CWS code (in standard
form) given by a graph $\mathcal{G}$ and word operators
$\{W_{l}=Z^{\mathbf{c}}\}_{\mathbf{c}\in \mathcal{C}}$ detects errors
in the set $\mathcal{E}$ if and only if the classical code
$\mathcal{C}$ detects errors from the set
$\clg(\mathcal{E)}$, and for each $E\in\mathcal{E}$,
\begin{eqnarray}
\text{either}\;\,\clg(E)& \neq& \mathbf{0}, \label{bei11}\\
\text{or, for each $i$,}\;\;\,
Z^{\mathbf{c}_{i}}E&=&EZ^{\mathbf{c}_{i}}. \label{bei12}
\end{eqnarray}
The code $\mathcal{Q}$ is non-degenerate (and also pure, see
Appendix~\ref{app:orthogonality}) iff condition~(\ref{bei11}) is satisfied for
all errors in $\mathcal{E}$ \cite{Cross-CWS-2009,Chuang-CWS-2009}.  For a
degenerate code, condition~(\ref{bei12}) needs to be ensured for errors that
do not satisfy Eq.~(\ref{bei11}).

The beauty of the CWS construction is that, for a given code in standard
form, we no longer need to worry about possible degeneracies.  The
classical error patterns induced by the function $\clg(\text{\boldmath
  $\cdot$})$ also separate the errors into corresponding degeneracy
classes \cite{Cross-CWS-2009,Chuang-CWS-2009}.

The CWS framework is general enough to also include all stabilizer
codes \cite{Cross-CWS-2009}.  Specifically, a stabilizer code
$[[n,k,d]]$ with the stabilizer $\mathscr{S}_0=\langle G_1,\ldots
G_{n-k}\rangle$ and logical operators ${\overline Z}_i$, ${\overline
  X}_i$, corresponds to a CWS code with the stabilizer
\begin{equation}
\mathscr{S}=\langle G_1,\ldots G_{n-k}, {\overline Z}_1,\ldots,
{\overline Z}_k\rangle\label{eq:stab1}
\end{equation}
and the codeword operator set
$\mathscr{W}=\langle{\overline X}_{1},\ldots,{\overline X}_{k}\rangle$
forming a group of size $K=2^k$.  Generally, an LC transformation is
required to obtain standard form of this code.

Conversely, an additive CWS code $\mathcal{Q}=((n,K,d))$ where the
codeword operators form an Abelian group (in which case $K=2^k$ with
integer $k$) is a stabilizer code $[[n,k,d]]$
 \cite{Cross-CWS-2009}.  In Sec.~\ref{sec:conversion} we show that the
$n-k$ generators $G_i$ of the stabilizer
can be obtained from the
graph-state generators $S_i$ [Eq.~(\ref{eq:canonical-generators})] by
a symplectic Gram-Schmidt orthogonalization procedure \cite{Wilde-2009}
which has no effect on the codeword operators.

\textbf{Example 3.} Consider a non-degenerate CWS code
$((5,6,2))$. The corresponding ring graph is illustrated in
Fig.~\ref{fig:tria}(b) (Ref.~\cite{Cross-CWS-2009}).  The $n=5$
generators of the stabilizer $\mathscr{S}$ are $S_{1}=XZIIZ$ and its
four cyclic permutations.  The corresponding stabilizer state has a
structure similar to Eq.~(\ref{eq:s3}), but with more terms.  Word
operators have the form $Z^{\mathbf{c}_{i}}$, with the classical
codewords
\begin{equation}
\begin{split}
\mathbf{c}_{0}  &  =00000,\quad   \mathbf{c}_{1}=01101,\quad
\mathbf{c}_{2}  =10110,\\  \mathbf{c}_{3} &=01011,\quad
\mathbf{c}_{4}   =10101,\quad  \mathbf{c}_{5}=11010.
\end{split}
\label{classicalcode}
\end{equation}
\strut\hfill$\Box$

\textbf{Example 4.} To express the $[[5,1,3]]$ stabilizer code [see
  Example 1] as a $((5,2,3))$ CWS code in standard form, we explicitly
construct alternative generators $S_i$ of the stabilizer
$\mathscr{S}=\langle G_{1},G_{2},G_{3},G_{4},\Z\rangle$
[Eq.~(\ref{eq:stab1})] to contain only one $X$ operator each.  We
obtain $S_3=G_1G_2\Z=IZXZI$ and its four cyclic permutations. This
does not require any qubit rotations due to a slightly unconventional
choice of the logical operators in Eq.~(\ref{eq:logical513}).  The
corresponding graph is the ring \cite{Cross-CWS-2009}, see
Fig.~\ref{fig:tria}(b).  The codeword operators are
$\mathscr{W}=\left\{ I,\X\right\} $, which by
Eq.~(\ref{eq:logical513}) correspond to classical binary codewords
$\{00000,11111\}$.  Note that the error map induced by the graph is
different from the mapping to group elements in Example 1; in
particular, $Z^{\clg(Z_2)}=Z_2$, $Z^{\clg(X_2)}=Z_1Z_3$,
$Z^{\clg(Y_2)}=Z_1Z_2Z_3$.\hfill $\Box$

\section{Generic recovery for CWS codes}
\label{sec:generic}

In this section we construct a generic recovery algorithm which can be
adapted to any non-additive code.  To our best knowledge, such an
algorithm has not been explicitly constructed before.

The basic idea is to construct a non-Pauli
measurement operator $M_\mathcal{Q}=2P_\mathcal{Q}-\openone$, where
$P_\mathcal{Q}\equiv\sum_{i=1}^{K}{|w_{i}\rangle\langle w_{i}|}$ is
the projector onto the code $\mathcal{Q}$ spanned by the orthonormal
basis $\{|w_i\rangle\}_{i=1}^{K}$.  The measurement operator is further
decomposed using the identity
\begin{equation}
-M_{\mathcal{Q}}=  \openone-2\sum_{i}^{K}|w_i\rangle\langle
    w_i|=\prod_{i}^{K}(\openone-2|w_{i}\rangle\langle w_{i}|).
\label{eq:product-identity}
\end{equation}

\begin{figure}[htbp]
  \centering
  \includegraphics[scale=1]{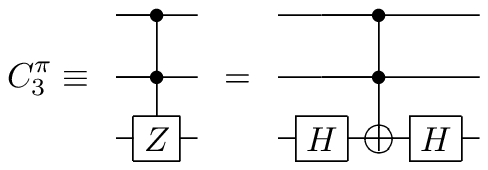}
  \caption{Decomposition of $n$-qubit controlled-$Z$ gate $C_n^Z$ in
    terms of $(n-1)$-controlled CNOT gate ($n$-qubit Toffoli gate) for $n=3$.}
  \label{fig:cZ-decomposition}
\end{figure}

We use the graph state encoding unitary $U_{\mathcal{G}}$ from
Eq.~(\ref{stabilizedstate}) and the following decomposition
\cite{Cross-CWS-2009} of the standard-form basis of the CWS code in
terms of the graph state $|s\rangle$ and the classical states
$|c_i\rangle$:
\[
|w_i\rangle
  =Z^{\mathbf{c}_{i}}|s\rangle
=Z^{\mathbf{c}_{i}}U_{\cal G}|0\rangle_n
=U_{\cal G}X^{\mathbf{c}_{i}}|0\rangle_n
=U_{\cal G}|\mathbf{c}_{i}\rangle.
\]
The measurement operator $M_\mathcal{Q}$ is rewritten as a
product
\begin{equation}
M_\mathcal{Q}=-U_{\cal G}M_\mathcal{C}
U_{\cal G}^{\dag}, \quad M_\mathcal{C}\equiv\prod_{i=1}^K M_{{\bf c}_i}
\label{eq:MirrorOperator}
\end{equation}
where the measurement operator
\begin{equation}
M_{{\bf c}_i}\equiv
\openone-2|\mathbf{c}_{i}\rangle\langle\mathbf{c}_{i}|
=
X^{\overline{\mathbf{c}}_{i}}\bigl(\openone-2|1\rangle_n\langle1|_n\bigr)
  X^{\overline{\mathbf{c}}_{i}},\label{eq:factorizationmirror1}
\end{equation}
stabilizes the orthogonal complement of the classical state $|{\bf
  c}_i\rangle$.  The components of the binary vector
${\overline{\mathbf{c}}_i}$ are the respective complements of ${\bf
  c}_i$, ${\overline{{c}}_{i,j}}=1\oplus {c}_{i,j}$.  The operator in
the parentheses in Eq.~(\ref{eq:factorizationmirror1}) is the
$(n-1)$-controlled $Z$ gate (the $Z$-operator is applied to the $n$-th
qubit only if all the remaining qubits are in state $|1\rangle$).
It can also be represented as $n$-qubit controlled phase gate
$C_n^\theta\equiv \exp(i\theta P_{|1\rangle_n})$ with $\theta=\pi$,
where the operator $P_{|1\rangle_n}$ projects onto the state
$|1\rangle_n$.  This can be further decomposed as a product of two
Hadamard gates and an $(n-1)$-controlled CNOT gate
[Fig.~\ref{fig:cZ-decomposition}] which for $n\ge 6$ can be
implemented in terms of one ancilla and $8(n-4)$ three-qubit Toffoli
gates \cite{Barenco-1995} and therefore has linear complexity in $n$.
With no ancillas, the complexity of the $(n-1)$-controlled CNOT gate
is $\mathcal{O}(n^2)$ \cite{Barenco-1995}.

\begin{figure*}[htbp]
  \centering
  \includegraphics[scale=1.0]{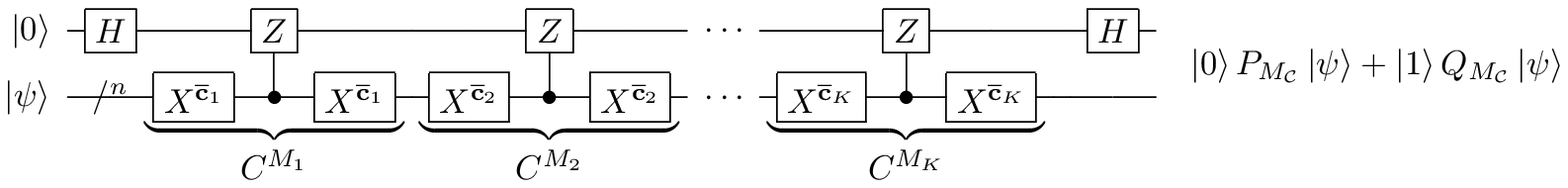}
  \caption{Generic measurement of the classical part $M_\mathcal{C}$ of the
    CWS code stabilizer $M_\mathcal{Q}$
    [Eq.~(\ref{eq:MirrorOperator})] uses $K$ $(n+1)$-qubit
    controlled-$Z$ gates.  Here $X^{{\overline {\bf c}}_i}$ indicates
    that a single-qubit gate $X^{{\overline {\bf c}}_{i,j}}$ is
    applied at the $j$\,th qubit, $j=1,\ldots n$, and $C^{M_i}$ is the
    controlled-$M_{{\bf c}_i}$ gate [see
      Eq.~(\ref{eq:factorizationmirror1})].  This can be further
    simplified by combining the neighboring $X^{\overline
      c}_{i,j}$-gates and replacing the controlled-$Z$ gates
    by $(n+1)$-qubit Toffoli gates as in
    Fig.~\protect\ref{fig:cZ-decomposition}.}
  \label{fig:gen}
\end{figure*}

The corresponding ancilla-based measurement for $M_{{\bf c}_i}$ can be
constructed with the help of two Hadamard gates
[Fig.~\ref{fig:measurementM}] by adding an extra control to each
$C_n^\pi$ gate.  Indeed, this correlates the state $|1\rangle$ of the
ancilla with $M_{{\bf c}_i}$ acting on the $n$ qubits, and the state
$|0\rangle$ of the ancilla with $\openone$.  When constructing the
measurement for the product of the operators $M_{{\bf c}_i}$, it is
sufficient to use only one ancilla, since for each basis state
$|w_i\rangle$ only one of these operators acts non-trivially.  The
classical part of the overall measurement circuit without the graph
state encoding $U_{\mathcal{G}}$ is shown in Fig.~\ref{fig:gen}.

The complexity of measuring $M_{\mathcal{Q}}$
[Eq.~(\ref{eq:MirrorOperator})] then becomes $K$ times the complexity
of $(n+1)$-qubit Toffoli gate for measuring each $M_{{\bf c}_i}$, plus
the complexity of the encoding circuit $U_{\mathcal{G}}$ and its
inverse $U_{\mathcal{G}}^\dagger$, which is at most $n^2$.
Overall, for large $n$, the measurement complexity is no larger than
$n^2+K \mathcal{O}(n)$, or $(1+K)\mathcal{O}(n^2)$ for a circuit
without additional ancillas.

We would like to emphasize that so far we have only constructed the
measurement for {\em error detection\/}.  Actual {\em error
  correction\/} for a non-additive code in this scheme involves
constructing measurements $M_{E}\equiv E M_\mathcal{Q} E^\dagger$ for
all corrupted subspaces corresponding to different degeneracy classes
given by different $\clg(E)$.  This relies on the orthogonality of the
corrupted subspaces, see Appendix \ref{app:orthogonality}.  For a
general $t$-error correcting code, the number of these measurements
can reach the same exponential order $B(n,t)$ as the number of
correctable errors in Eq.~(\ref{eq:sphere}).  For non-degenerate
codes, we cannot do better using this method.

Note that the measurement circuit derived in this section first
decodes the quantum information, then performs the measurement for the
classical code, and finally re-encodes the quantum state.

\begin{figure*}[htbp]
  \centering
  \includegraphics[scale=1.0]{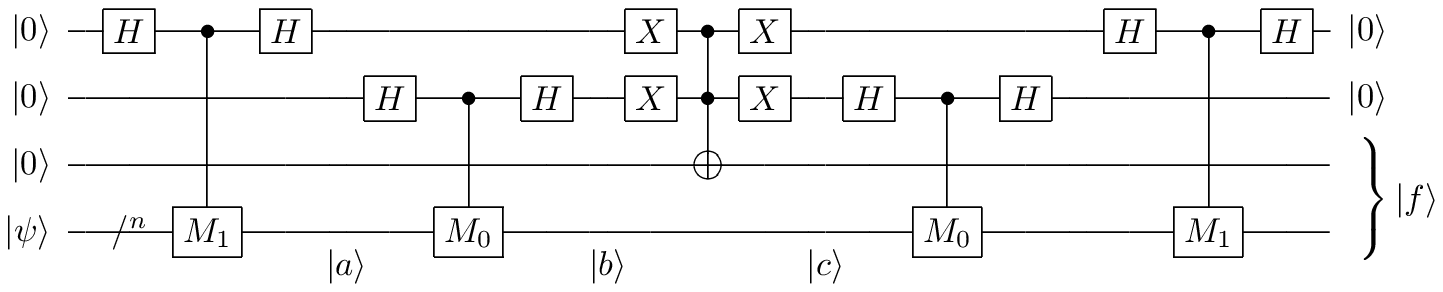}
  \caption{Measurement
    $M_{1}\wedge M_{0}$ by performing logical AND operation on two
    ancillas.  We use the notations $P_i\equiv P_{M_i}$, $Q_i\equiv
    \openone-P_i$, $i=0,1$ and assume $P_1P_0=P_0P_1$.  The circuit
    returns $\ket f=\ket{1} P_{1} P_{0} \ket\psi+\ket0(\openone-P_{1}
    P_{0})\ket\psi$.  Intermediate results are: $\ket
    a=\ket{000}P_{1}\ket\psi+\ket{100}Q_{1}\ket\psi$, $\ket
    b=\ket{000}P_{1} P_{0}\ket\psi+\ket{010}P_{1} Q_{0}\ket\psi
    +\ket{100}Q_{1} P_{0}\ket\psi+ \ket{110}Q_{1} Q_{0}\ket\psi$,
    $\ket c=\ket{001} P_{1}P_{0}\ket\psi+\ket{010}P_{1} Q_{0}\ket\psi+
    \ket{100}Q_{1} P_{0}\ket\psi+ \ket{110}Q_{1}Q_{0}\ket\psi$; the
    last two groups of gates return the first two ancillas to the state
    $\ket{00}$.}
\label{measurementAnd}
\end{figure*}

\begin{figure*}[htbp]
\centering
\includegraphics[scale=1.0]{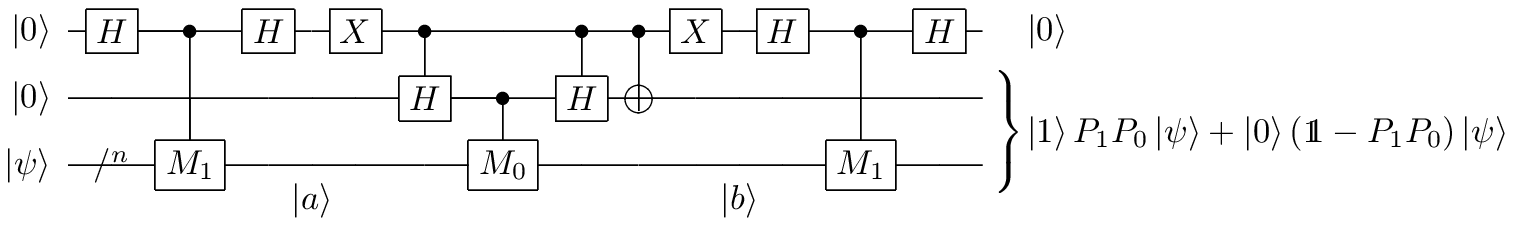}
\caption{Simplified measurement for
  $M_{1}\wedge M_{0}$; notations as in Fig.~\ref{measurementAnd}.
  Intermediate results (cf.\
  Fig.~\ref{fig:measurementM}) are:
  $\ket{a}=\ket{00}{P}_1\ket\psi+\ket{10}{Q}_1\ket\psi$,
  $\ket{b}=\ket0(\ket{1}{P}_0+\ket{0}{Q}_0)
  {P}_1\ket\psi+\ket{10}{Q}_1\ket\psi$.  The effect of the last
  block is to select the component with the first ancilla in the state
  $\ket0$; this requires that the projectors ${P}_1$ and
  ${P}_0$ commute.  The result is equivalent to
  $\ket1{P}_{M_1\wedge
    M_0}\ket\psi+\ket0(\openone-{P}_{M_1\wedge M_0})\ket\psi$. }
\label{measurementAnd_new}
\end{figure*}

\begin{figure*}[ptbh]
  \centering\includegraphics[scale=1.0]{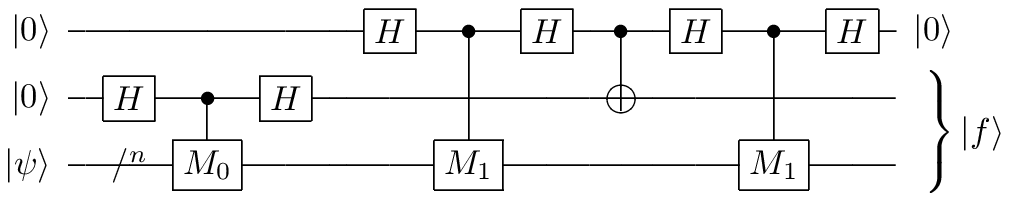}
  \caption{Measurement
    $M_{1}\bplus M_{0}$ by performing logical XOR gate on the two
    ancillas and subsequent recovery of the first ancilla.  Notations as
    in Fig.~\ref{measurementAnd}.  The final result
    is $\ket f=\ket1 (P_0 Q_1+P_1 Q_0)\ket\psi+ \ket0 (P_0
    P_1+Q_0Q_1)\ket\psi$. }
\label{measurementXor}
\end{figure*}
\begin{figure}[htbp]
  \centering
  \includegraphics[scale=1.0]{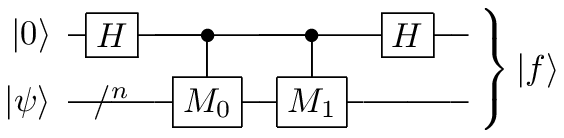}
  \caption{Simplified measurement for $M_1\bplus M_0$.  Notations as
    in Fig.~\ref{measurementAnd}.  The result  (cf.\
  Fig.~\ref{fig:measurementM})
  $\ket f=\ket1({Q}_1{P}_0+{P}_1{Q}_0)
  \ket\psi
  +\ket0({P}_1{P}_0+{Q}_1{Q}_0)\ket\psi$
  is equivalent to $\ket1{P}_{M_1\bplus
    M_0}\ket\psi+\ket 0(\openone-{P}_{M_1\bplus M_0})\ket\psi$.}
\label{fig:measurementM1plusM0}
\end{figure}

\section{Measurements for USt codes}\label{sec:ust-measurement}

In this section we construct a quantum circuit for the measurement operator
${M}_\mathcal{Q}$ of a USt code $((n,K 2^k, d))$.  To this end, we define the
logical combinations of non-Pauli measurements in agreement with analogous
combinations defined in Ref.~\cite{Aggarwal-Calderbank-2008} for the
projection operators, and construct the circuits for logical combinations
{AND} [Figs.~\ref{measurementAnd}, \ref{measurementAnd_new}] and {XOR}
[Figs.~\ref{measurementXor}, \ref{fig:measurementM1plusM0}].  We use these
circuits to construct the measurement for $M_\mathcal{Q}$ with complexity not
exceeding $2 K (n+1)(n-k)$.

\subsection{Algebra of measurements}\label{sec:logic}
{\em Logical AND\/}: Given two commuting measurement operators $M_{1}$
and $M_{0}$, let $M_{1}\wedge M_{0}$ denote the measurement operator
that stabilizes all states in the subspace
\begin{equation}
  \mathcal{P}(M_{1}\wedge
  M_{0})\equiv\mathcal{P}(M_{1})\cap\mathcal{P}(M_{0}).
  \label{eq:space-intersection}
\end{equation}
The output of the measurement $M_{1}\wedge M_{0}$ is identical to the logical
{AND} operation performed on the output of measurements $M_{1}$ and $M_{0}$.
This measurement can be implemented by the circuit in
Fig.~\ref{measurementAnd}.  Here the first two ancillas are entangled with the
two measurement outcomes; the third ancilla is flipped only if both ancillas
are in the $\ket0$ state, which gives the combination $\ket 1 P_{M_1\wedge
  M_0}\ket\psi$.

\begin{figure}[ptbh]
\centering\includegraphics[scale=1]{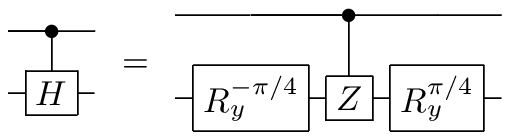}
\caption{Implementation of the controlled-$H$
  gate based on the identity $H=\exp(-i Y \pi/8) Z \exp(i Y \pi/8)$.}
\label{CHgate}
\end{figure}

The projector onto the positive eigenspace of $M_1\wedge M_0$ satisfies
the identity
\begin{equation}
  P_{M_1\wedge M_0}=P_{M_1} P_{M_0}.  \label{eq:product-identity-AND}
\end{equation}
This identity can be used to obtain a simplified circuit which
only uses two ancillas, see Fig.~\ref{measurementAnd_new}, with the
price of two additional controlled-Hadamard gates
[Fig.~\ref{CHgate}].

The circuits in Figs.~\ref{measurementAnd} and
\ref{measurementAnd_new} can be generalized to perform the measurement
corresponding to the logical AND of $\ell> 1$ commuting measurement operators
with the help of associativity, e.g., $M_{2}\wedge M_1 \wedge M_0= 
M_2\wedge (M_1\wedge M_0)$.  The generalization of the simplified
circuit in Fig.~\ref{measurementAnd_new} requires only two ancillas
for any $\ell>1$.  The corresponding complexity is $2\ell-1$ times the
complexity of a controlled-$M$ gate, plus $2\ell-1$ times the
complexity of a controlled single-qubit gate.  When all $M_i$ are
$n$-qubit Pauli operators, the overall complexity with two ancillas is
$(2\ell-1)(n+1)$.

{\em Logical XOR\/}: In analogy to the logical ``{exclusive OR}'', we
define the symmetric difference $A\bigtriangleup B\bigtriangleup C\ldots$ of
vector spaces $A$, $B$, $C$, \ldots\ as the vector space formed by the
basis vectors that belong to an odd number of the original vector
spaces.  For two vector spaces
\begin{equation}
  A\bigtriangleup B\equiv
  (A\cap B^\perp)\oplus(B\cap A^\perp).\label{eq:symmetric-difference}
\end{equation}
This operation is obviously associative, $A\bigtriangleup
(B\bigtriangleup C)= (A\bigtriangleup B)\bigtriangleup C$.  For two
commuting measurement operators $M_0$, $M_1$, let $M_1\bplus M_0$ be
the measurement operator that stabilizes the subspace
$\mathcal{P}(M_1)\bigtriangleup\mathcal{P}(M_0)$. Explicitly,
\begin{eqnarray}
  \nonumber
  \lefteqn{\mathcal{P}(M_1\bplus M_0)\equiv
    \mathcal{P}(M_{1})\bigtriangleup\mathcal{P}(M_{0})} & & \\
  &=&
  \left[\mathcal{P}(M_{1})\cap\mathcal{P}^\perp(M_{0})\right]
  \oplus
  \left[\mathcal{P}(M_{0})\cap\mathcal{P}^\perp(M_{1})\right].
  \label{eq:oplus-space}
\end{eqnarray}
The output of measuring $M_{1}\bplus M_{0}$ is identical to the
logical {XOR} operation performed on the outputs of measurements
$M_{1}$ and $M_{0}$.
The corresponding  measurement
can be implemented by combining the two ancillas
with a CNOT gate [Fig.~\ref{measurementXor}].

To simplify this measurement, we show that $M_1\bplus
M_0=-M_1M_0$.  Indeed, Eq.~(\ref{eq:oplus-space}) implies that for the
projection operators $P_i\equiv P_{M_i}$, $i=0,1$,
$$
P_{M_1\bplus M_0}=P_1(\openone -P_0)+P_0(\openone-P_1).
$$
The corresponding measurement operator is factorized with the help
of the projector identities~(\ref{eq:projector-identities}),
\begin{eqnarray*}
{M_1\bplus M_0}&=&2[P_1(\openone -P_0)+P_0(\openone-P_1)]-\openone\\
&=&-(2P_1-\openone )(2P_0-\openone )=-M_1M_0.
\end{eqnarray*}
This implies that $\mathcal{P}(M_1\bplus M_0)=\mathcal{P}^\perp(M_1
M_0)$.  In other words, the measurement of $M_1\bplus M_0$ can be
implemented simply as an (inverted) concatenation of two measurements,
see Fig.~\ref{fig:measurementM1plusM0}.  The same circuit can also be
obtained from that in Fig.~\ref{measurementXor} by a sequence of
circuit simplifications (not shown).

The circuit in Fig.~\ref{fig:measurementM1plusM0} is immediately
generalized to a combination of more than two measurements, $M_1\bplus
\ldots \bplus M_{\ell}=(-1)^{l-1}M_1\ldots M_{\ell}$.  The
corresponding complexity for computing the XOR of $\ell$ measurements
is simply the sum of the individual complexities, implying that this
concatenation has no overhead.

\subsection{Error detection for union stabilizer codes}
A USt code $\mathcal{Q}=((n,K\,2^k,d))$ is a direct sum
(\ref{eq:ust-defined}) of $K$ mutually orthogonal subspaces obtained
by translating the originating stabilizer code
$\mathcal{Q}_0=[[n,k,d_0]]$.  For mutually orthogonal subspaces
$A\perp B$, we have $A\subset B^\perp$ and $B\subset A^\perp$, and the
direct sum is the same as the symmetric difference
(\ref{eq:symmetric-difference}), $A\oplus B=A\bigtriangleup B$.

In turn, the stabilizer code $\mathcal{Q}_0$ is an intersection of the
subspaces stabilized by the generators $G_i$ of the stabilizer, see
Eq.~(\ref{eq:stabilizer-code-intersection}).  The translated subspaces
$t_j(\mathcal{Q}_0)$ are stabilized by the Pauli operators
$M_{i,j}=t_j G_i t_j^\dagger$.  We can therefore decompose the USt
code $\mathcal{Q}$ as
\begin{equation}
  \mathcal{Q} =
  \bigoplus_{j=1}^{K}\biggl[  \bigcap_{i=1}
  ^{n-k}{\mathcal{P}(M_{i,j})}\biggr]  =
  \mathop{\text{\raisebox{-0.75ex}{\LARGE{$\bigtriangleup$}}}}
  \limits_{j=1}^{K}\biggl[  \bigcap_{i=1}
  ^{n-k}{\mathcal{P}(M_{i,j})}\biggr]  .
  \label{eq:ust-space-decomposition}
\end{equation}
This gives the decomposition of the measurement operator
$M_{\mathcal{Q}}$ whose positive eigenspace is the code $\mathcal{Q}$
as
\begin{equation}
  M_{\mathcal{Q}}=\bigbplus_{j=1}^{K}\left[  {\bigwedge_{i=1}^{n-k}{M_{i,j}}}\right].
  \label{eq:ust-decomposition}
\end{equation}

Recall that the complexity of each of the $K$ logical AND operations
is $[2(n-k)-1](n+1)$.  No additional overhead is required to form the
logical XOR of the results.  Thus, we obtain the following
\begin{theorem}
\label{theorem:complexity1}
Error detection for a USt code of length $n$ and
dimension $K \,2^{k}$, formed by a translation set of size $K>0$, has
complexity at most $2K(n-k)(n+1)$.
\end{theorem}

Note that in the special case of a CWS code ($k=0$), the prefactor of
$K$ is quadratic in $n$ whereas the corresponding prefactor obtained
in Sec.~\ref{sec:generic} is {\em linear\/} in $n$.  The reason is
that in Eq.~(\ref{eq:MirrorOperator}) the graph encoding circuit
${Q}_\mathcal{G}$ with complexity $\mathcal{O}(n^2)$ is used only
twice, and the projections onto the classical states have linear
complexity.  In Eq.~(\ref{eq:ust-decomposition}) we are using $K$
projections onto basis states of the quantum code.  The advantage of
the more complex measurement constructed in this section is that it
does not involve having unprotected decoded qubits for the entire
duration of the measurement.

\section{Structured measurement for CWS codes}
\label{sec:clusters}

\subsection{Grouping correctable errors}\label{sec:grouping}
Recall from Section~\ref{sec:stabilizer-codes} that for stabilizer
codes the representatives of the error degeneracy classes form an Abelian
group whose generators are in one-to-one correspondence with the
generators of the stabilizer.  Measuring the $n-k$ generators
of the stabilizer of a stabilizer code $[[n,k,d]]$ uniquely identifies
the degeneracy class of the error.

In this section we establish a similar structure for CWS codes.
First, for any subset $\mathcal{D}\subset\mathcal{E}$ of correctable
errors of a quantum code $\mathcal{Q}$, we define the set
$\mathcal{E}_{\overline{\mathcal{D}}}$ of {\em unrelated\/} errors which do
not fall in the same degeneracy class with any error from
$\mathcal{D}$.  The formal definition
\begin{equation}
\mathcal{E}_{\overline{\mathcal{D}}}\equiv \{E: E\in\mathcal{E}
\wedge C_{E,E_1}=0 ,\, \forall E_1\in\mathcal{D}\}
\label{eq:unrelated}
\end{equation}
is based on the general error correction
condition~(\ref{SNcondtion-correcting}) and the orthogonality of corrupted
spaces, see Appendix~\ref{app:orthogonality}.  When errors in $\mathcal{E}$
are non-degenerate, the definition~(\ref{eq:unrelated}) is equivalent to the
set difference, $\mathcal{E}\setminus\mathcal{D}$.  In the general case, since
we do not distinguish between mutually degenerate errors,
$\mathcal{E}_{\overline{\mathcal{D}}}$ can be thought of as the difference
between the sets of degeneracy classes in $\mathcal{E}$ and in $\mathcal{D}$.

Definition (\ref{eq:unrelated}) implies that the subspaces
\begin{alignat*}{5}
\mathcal{D}(\mathcal{Q})\equiv \bigoplus_{E\in\mathcal{D}} E(\mathcal{Q})
\end{alignat*}
and $\mathcal{E}_{\overline{\mathcal{D}}}(\mathcal{Q})$, defined
analogously, are mutually orthogonal.  Moreover, if the elements of
the set $\mathcal{D}$ form a group $\mathscr{D}\equiv \mathcal{D}$,
the subspace $\mathscr{D}(\mathcal{Q})$ is also orthogonal to
$\mathcal{E}_{\overline {\mathscr{D}}}[\mathscr{D}(\mathcal{Q})]$ [see
  Eq.~(\ref{eq:err-detection-aux}) below].  In other words,
$\mathcal{Q}_\mathscr{D}\equiv \mathscr{D}(\mathcal{Q})$ can be viewed
as a quantum code which detects errors from
$\mathscr{E}_{\overline{\mathscr{D}}}$.

This observation, together with the error-detection measurement for USt
codes constructed in the previous section, forms the basis of our error
grouping technique.  We prove the following

\begin{theorem}
  For a CWS code $\mathcal{Q}=(\mathcal{G},\mathcal{C})$ in standard
  form and a group $\mathscr{D}$ formed by graph images of some
  correctable errors in $\mathcal{E}$, the code
  $\mathcal{Q}_\mathscr{D}\equiv \mathscr{D}(\mathcal{Q})$ is a USt
  code which detects all errors in
  $\mathcal{E}_{\overline{\mathscr{D}}}$.
  \label{theorem:error-detect}
\end{theorem}
\noindent\textbf{Proof}.  First, we show that the subspace
$\mathscr{D}(\mathcal{Q})$ is a USt code.  The corresponding set of
basis vectors is
\begin{equation}
  \label{eq:aux-basis}
    \mathscr{D}\left(\left\{\ket{w_1},\ldots,\ket{w_K}\right\}\right)
    \equiv \bigcup_{e_\alpha\in\mathscr{D}}
    \bigcup_{i=1}^K \left\{ e_\alpha \ket{w_i}\right\}.
\end{equation}
 These vectors are mutually orthogonal,
\begin{equation}
  \bra{w_i} e_\beta^\dagger e_\alpha\ket{w_j}=0,\;\, \forall i,j\le K,\;\,
  e_\alpha,e_\beta\in\mathscr{D}, \;\,e_\alpha\neq
  e_\beta,\label{eq:orthogonality}
\end{equation}
since every element $e_\alpha=Z^{\clg(E_\alpha)}$ of the group
$\mathscr{D}$ is a representative of a separate error degeneracy
class.  Further, the group $\mathscr{D}$ is Abelian, and its elements
commute with the codeword generators $W_i=Z^{\bf c}_i$,
$c_i\in\mathcal{C}$.  Therefore, using Eq.~(\ref{eq:codeword-def}), we
can rearrange the set~(\ref{eq:aux-basis}) as
\begin{equation}
  \label{eq:aux-basis-two}
    \mathscr{D}\left(\left\{\ket{w_1},\ldots,\ket{w_K}\right\}\right)
    = \bigcup_{i=1}^K W_i\left(   \left\{
    e_\alpha\ket{s}\right\}_{e_\alpha\in \mathscr{D}}\right).
\end{equation}
The set in the parentheses on the right hand side is a basis of the
additive CWS code $\mathcal{Q}_{0\mathscr{D}}$ formed by the group
$\mathscr{D}$ acting on the graph state $\ket s$.  Then, we can write
the subspace $\mathscr{D}(\mathcal{Q})$ explicitly as a USt code
[cf.~Eq.~(\ref{eq:ust-defined})]
\begin{equation}
  \label{eq:aux-code}
  \mathscr{D}(\mathcal{Q})=\bigoplus_{i=1}^K
  W_i(\mathcal{Q}_{0\mathscr{D}}),
\end{equation}
where the translations are given by the set of codeword operators
$\mathcal{W}\equiv \{W_i\}_{i=1}^K$ of the original code
$\mathcal{Q}$.  Orthogonality condition~(\ref{eq:ust-orthogonality})
is ensured by Eq.~(\ref{eq:orthogonality}).

Second, we check the error-detection
condition~(\ref{SNcondtion-detecting}) for the
code~(\ref{eq:aux-code}).  Explicitly, for an error $E\in
\mathcal{E}_{\overline{\mathscr{D}}}$, and for the orthogonal basis
states $e_\alpha W_i\ket s$,
\begin{equation}
  \label{eq:err-detection-aux}
  \bra s {W_i}^\dagger e_\alpha^\dagger  E e_\beta W_j \ket s
  =\pm   \bra{w_i}   E e_\alpha^\dagger e_\beta \ket{w_j}=0
\end{equation}
for all $\alpha,\beta,i,j$, according to
Eqs.~(\ref{SNcondtion-correcting}), (\ref{eq:unrelated})
and the group property of $\mathscr{D}$. \hfill $\Box$

Now, to correct errors in groups, we just have to find a suitable
decomposition of the graph images of the original error set into a
collection of groups, $\clg(\mathcal{E})=\bigcup_j\mathscr{D}_j$, and
perform individual error-detection measurements for the auxiliary
codes $\mathcal{Q}_{\mathscr{D}_j}$ until the group containing the
error is identified.

To find an error within a group $\mathscr{D}\equiv \langle g_1,\ldots
,g_m\rangle$ with $m$ generators, we can try all $m$ subgroups of
$\mathscr{D}$ with one generator missing.  More specifically, for a
generator $g_l$ we consider the subgroup $\mathscr{D}^{(l)}=\langle
g_1,\ldots,g_{l-1},g_{l+1},\ldots,g_m\rangle$ and perform error
detection for the code $\mathcal{Q}_{\mathscr{D}}^{(l)}\equiv
\mathscr{D}^{(l)}(\mathcal{Q})$.  After completing $m$ measurements,
we obtain a representative of the actual error class.  This is the
product of all generators $g_l$ for which the corresponding code
$\mathcal{Q}_{\mathscr{D}}^{(l)}$ detected an error.

\subsection{Complexity of a combined measurement}\label{sec:conversion}
To actually carry out the discussed program, we need to construct the
$n-m$ generators $G_i$ of the stabilizer of the code
$\mathcal{Q}_{0\mathscr{D}}$.  The generators have to commute with the
$m$ generators $g_\alpha$ in the group $\mathscr{D}$.

This can be done with the Gram-Schmidt (GS)
orthogonalization \cite{Wilde-2009} of the graph-state generators
$S_i$ [Eq.~(\ref{eq:canonical-generators})] with respect to the
generators $g_\alpha$.  As a result, we obtain the orthogonalized set
of independent generators $S_i'$ such that $g_\alpha S_i'
=(-1)^{\delta_{i\alpha}} S_i' g_\alpha$.  We can take the last $n-m$
of the obtained generators as the generators of the stabilizer,
$G_i=S'_{i+m}$, $i=1,\ldots, n-m$.

The orthogonalization procedure is guaranteed to produce exactly $m$
generators $S_\alpha'$ anti-commuting with the corresponding errors
$g_\alpha$, $\alpha=1,\ldots, m$.  Indeed, the GS orthogonalization
procedure can be viewed as a sequence of row operations applied to the
original $n\times m$ binary matrix $B$ with the elements $b_{i\alpha}$
which define the original commutation relation,
\begin{equation}
  S_{i}g_{\alpha}=(-1)^{b_{i\alpha}}g_{\alpha}S_{i}.
  \label{eq:commut}
\end{equation}
The generator $g_{\alpha}$ anti-commutes with at least one operator in
$\mathscr{S}$ if and only if the $\alpha$-th column of $B$ is not an
all-zero column.  Then all $m$ generators are independent (no
generator can be expressed as a product of some others) if and only if
$B$ has full column rank.

By this explicit construction, the generators $G_i$ of the stabilizer
of the auxiliary code $\mathcal{Q}_\mathscr{D}$
[Eq.~(\ref{eq:aux-code})] are Pauli operators in the original
graph-state basis.  The complexity of each error-detection measurement
$M({\mathcal{Q}_{\mathscr{D}}})$ is therefore given by Theorem
\ref{theorem:complexity1}.

\subsection{Additive CWS codes}\label{sec:additive}
The procedure described above appears to be extremely tedious, much
more complicated than the syndrome measurement for a stabilizer code.
However, it turns out that for stabilizer codes this is no more
difficult than the regular syndrome-based error correction.

Indeed, for a stabilizer code $\mathcal{Q}=[[n,k,d]]$, the degeneracy
classes for all correctable errors form a group of all translations of
the code, $\mathscr{D}\equiv
\mathscr{T}=\langle g_1,\ldots ,g_{n-k}\rangle$, with $n-k$ generators.
To locate the error, we just have to
go over all $n-k$ USt codes $\mathscr{D}^{(l)}(\mathcal{Q})$ generated
by the subgroups of $\mathscr{D}$ with the generator $g_l$ missing.
Since the originating code $\mathcal{Q}$ is a stabilizer code, the USt
codes $\mathscr{D}^{(l)}(\mathcal{Q})$ are actually stabilizer codes,
encoding $\tilde k=k+(n-k-1)=n-1$ qubits each.  Hence there is only
one non-trivial error, and up to error degeneracies,
$\mathcal{E}_{{\mathscr{D}}^{(l)}}=\langle g_l\rangle$.  The
corresponding stabilizers $\mathscr{S}^{(l)}$ have only one generator
each.  The necessary measurements are just independent measurements of
$n-k$ Pauli operators, the same as needed to measure the syndrome.
Moreover, if the error representatives $g_l$, $l=1,\ldots,n-k$, are
chosen to satisfy the orthogonality condition $G_i
g_l=(-1)^{\delta_{il}}g_l G_i$ as in Example 1, the operators to be
measured are the original generators $G_i$ of the stabilizer, and the
corresponding measurement is just the syndrome measurement.

\textbf{Example 5.}
Consider the additive code $((5,2,3))$ equivalent to the stabilizer
code $[[5,1,3]]$, see Examples 1 and 4.  The graph-induced maps of
single-qubit errors form a group of translations of the code,
$\mathscr{T}=\langle Z_1,Z_2,Z_3,Z_4\rangle$.  This group contains all
error degeneracy classes, $\mathscr{D}=\mathscr{T}$.  With the
addition of the logical operator $\overline X\equiv ZZZZZ$, these can
generate the entire $5$-qubit Hilbert space $\H5$ from the graph state
$\ket s$; we have $\mathscr{T}(\mathcal{Q})=\H5$.

Indeed,  if we form  a measurement as for a generic CWS code, we
first obtain the stabilizer of the auxiliary code $\mathcal{Q}_{0\mathscr{D}}$
[Eq.~(\ref{eq:aux-code})] which in this case has only one generator,
$\mathscr{S}=\langle S_5\rangle$, where $S_5=ZIIZX$, see Example 4.
Translating this code with the set (in this case, group)
$\mathscr{W}=\{I,\overline X\}$ of codeword operators, we get the auxiliary USt
code $\mathscr{W}(\mathcal{Q}_{0\mathscr{D}})$ as the union of the positive
eigenspaces of the operators  [see Eq.~(\ref{eq:ust-space-decomposition})],
$$M_{1,0}=S_5,\quad  M_{1,1}=\overline X S_5 \overline
X^\dagger =-S_5,
$$ which is the
entire Hilbert space,
$\mathscr{W}(\mathcal{Q}_{0\mathscr{D}})=\mathscr{D}(\mathcal{Q})=\H5$, as
expected.

To locate the error within the group $\mathscr{D}$ with $m=4$
generators, we form a set of smaller codes
$\mathcal{Q}_{\mathscr{D}}^{(l)}\equiv
\mathscr{D}^{(l)}(\mathcal{Q})$, $l=1,\ldots, 4$, where the group
$\mathscr{D}^{(l)}$ is obtained from $\mathscr{D}$ by removing the
$l$-th generator.  The corresponding stabilizers are
$\mathscr{S}^{(1)}=\langle S_1,S_5\rangle$, $\mathscr{S}^{(2)}=\langle
S_2,S_5\rangle$, etc.  The matrices $M_{i,j}^{(l)}$ of conjugated
generators have the form, e.g.,
$$
M_{i,j}^{(1)}=\left(
  \begin{array}[c]{cc}
    S_1,& -S_1\\
    S_5,& -S_5
  \end{array}
\right), \quad
M_{i,j}^{(2)}=\left(
  \begin{array}[c]{cc}
    S_2,& -S_2\\
    S_5,& -S_5
  \end{array}
\right), \ldots
$$
The code $\mathcal{Q}_{\mathscr{D}}^{(l)}$ is formed as the union of
the common positive eigenspaces of the operators in the columns of the
matrix $M_{i,j}^{(l)}$.  Clearly, these codes can be more compactly
introduced as positive eigenspaces of the operators $\tilde G_l=S_l
S_5$, $l=1,\ldots, 4$.  Such a simplification only happens when the
original code $\mathcal{Q}$ is additive.  While the operators $\tilde
G_i$ are different from the stabilizer generators in
Eq.~(\ref{eq:stab513}), they generate the same stabilizer
$\mathscr{S}=\langle \tilde G_1,\ldots,\tilde G_4\rangle$ of the
original code $\mathcal{Q}$.  It is also easy to check that the same
procedure gives the original generators $G_i$ [Eq.~(\ref{eq:stab513})]
if we start with the error representatives (\ref{eq:513-errors}).
\hfill $\Box $

\subsection{Generic CWS codes}\label{sec:generic-cws}
Now consider the case of a generic CWS code $\mathcal{Q}=((n,K,d))$.
Without analyzing the graph structure, it is impossible to tell
whether there is any set of classical images of correctable errors
that forms a large group.  However, since we know its minimum
distance, we know that the code can correct errors located on
$t=\lfloor(d-1)/2\rfloor$ qubits.  All errors located on a given set
of qubits form a group.  Therefore, by taking an \emph{index set}
$A\subset\{1,\ldots, n\}$ of $s\le t$ different qubit positions, we
can ensure that the corresponding correctable errors
$\{E_j\}_{j=1}^{4^s}$ form a group with $2s$ independent generators.
The corresponding graph images $\clg(E_{j})\in\mathscr{D}_A$ obey the
same multiplication table, but they are not necessarily independent.
As a result, the Abelian group $\mathscr{D}_A$ generally has $m\le 2s$
generators.  Since all group elements correspond to correctable
errors, the conditions of Theorem~\ref{theorem:error-detect} are
satisfied.

Overall, to locate an error of weight $t$ or less, we need to
iterate over each (but the last one) of the $\binom{n}{t}$ index sets
of size $t$ and perform the error-detecting measurements in the
corresponding USt codes $\mathcal{E}_A(\mathcal{Q})$ until the index
set with the error is found.  This requires up to $\binom{n}{t}-1$
measurements to locate the index set, and the error can be identified
after additional $m\le 2t$ measurements.  This can be summarized as
the following
\begin{theorem}
  A CWS code of distance $d$ can correct errors of weight up to
  $t=\left\lfloor (d-1)/2\right\rfloor$ by performing at most
  \begin{equation}
    N(n,t)\equiv {\binom{n}{t}}+2t-1 \label{eq:measur-new}
  \end{equation}
measurements.
\end{theorem}
For any length $n\geq3$, this scheme reduces the total
number~(\ref{eq:sphere}) of error patterns  by a factor
\begin{equation}
  \frac{B(n,t)}{N(n,t)}\geq\left\{
    \begin{array}[c]{ll}
      \displaystyle
      \frac{3n+1}{n+1}, & \text{if}\;\,t=1,\\
      \displaystyle
      \;3^{t^{\text{\strut}}}, & \text{if}\;\,t>1.
    \end{array}
  \right.  \label{factoreven}
\end{equation}

\textbf{Example 6.} Consider the $((5,6,2))$ code previously discussed
in Example 3.  While the distance $d=2$ is too small to correct
arbitrary errors, we can correct an error located at a given qubit.
Assume that an error may have happened on the second qubit.  Then we
only need to check the index set $A=\{2\}$.  The errors
$\{\openone,X_2,Y_2,Z_2\}$ located in $A$ form a group with generators
$\{X_2,Z_2\}$; the corresponding group of classical error patterns
induced by the ring graph in Fig.~\ref{fig:tria}(b) is
$\mathscr{D}_A=\langle Z_1Z_3,Z_2\rangle$.  The three generators $G_i$
of the stabilizer of the originating USt code
$\mathcal{Q}_{0\mathscr{D}_A}$ can be chosen as, e.g.,
$G_1=S_1S_3=XIXZZ$, $G_2=S_4=IIZXZ$, $G_3=S_5=ZIIZX$.  Using the
classical codewords~(\ref{classicalcode}) for the translation
operators $t_j=Z^{{\bf c}_j}$, we obtain the conjugated generators
$M_{i,j}=t_j G_i t_j^\dagger$
\begin{equation}
  \label{eq:translated-562}
  M_{i,j}=\left(
    \begin{array}[c]{rrrrrr}
      G_1, &-G_1, & G_1, & G_1, & G_1, &-G_1\\
      G_2, & G_2, &-G_2, &-G_2, & G_2, &-G_2\\
      G_3, &-G_3, & G_3, &-G_3, &-G_3, & G_3
\end{array}\right).
\end{equation}
According to Eq.~(\ref{eq:ust-space-decomposition}), the auxiliary
code $\mathcal{Q}_{\mathscr{D}_A}$ is
 a direct
sum of the common positive
eigenspaces of the operators in the six columns of the
matrix~(\ref{eq:translated-562}).

To locate the actual error in this $24$-dimensional space, we consider
the two subgroups $\mathscr{D}^{(1)}=\langle ZIZII\rangle$ and
$\mathscr{D}^{(2)}=\langle IZIII\rangle$ of $\mathscr{D}_A$.  The
stabilizers of the corresponding auxiliary codes
$\mathcal{Q}_{\mathscr{D}_A}^{(1)}$ and
$\mathcal{Q}_{\mathscr{D}_A}^{(2)}$ can be obtained by adding
$G_4^{(1)}=S_2=ZXZII$ and $G_4^{(2)}=S_3=IZXZI$, respectively; this adds
one of the rows
\begin{alignat}{10}
  \label{eq:row1}
  M_{4,j}^{(1)}&{}=(S_2,-S_2,&S_2, &&-S_2,&&S_2, &&-S_2&),\\
  M_{4,j}^{(2)}&{}=(S_3,-S_3,&-S_3,&& S_3,&&-S_3,&&S_3&)
\end{alignat}
to the matrix~(\ref{eq:translated-562}).  The original code
$\mathcal{Q}$ is the intersection of the codes
$\mathcal{Q}_{\mathscr{D}_A}^{(1)}$ and
$\mathcal{Q}_{\mathscr{D}_A}^{(2)}$; the corrupted space
$X_2(\mathcal{Q})$ is located in $\mathcal{Q}_{\mathscr{D}_A}^{(1)}$,
but not in $\mathcal{Q}_{\mathscr{D}_A}^{(2)}$, while, e.g., the
corrupted space $Y_2(\mathcal{Q})$ is located in
$\mathcal{Q}_{\mathscr{D}_A}$, but not in
$\mathcal{Q}_{\mathscr{D}_A}^{(1)}$ or
$\mathcal{Q}_{\mathscr{D}_A}^{(2)}$. \nolinebreak\hfill $\Box $

\subsection{General USt codes}
A similar procedure can be carried over for a general USt code
$((n,K\,2^k,d))$, with the only difference that the definitions of the
groups $\mathscr{D}$ and the auxiliary codes
$\mathcal{Q}_{0\mathscr{D}}$ [Eq.~(\ref{eq:aux-code})] should also
include the $k$ generators of the originating stabilizer code
$\mathcal{Q}_0$ [Sec.~\ref{sec:ust}].  Overall, the complexity of
error recovery for a generic USt code can be summarized by the
following
\begin{theorem}
\label{main}Consider any $t$-error correcting USt code of
length $n$ and dimension $K\,2^{k}$, with the translation set of size
$K$.  Then this code can correct errors using ${\binom{n}{t}}+2t-1$ or
fewer measurements, each of which has complexity $2K (n+1)(n-k-1)$
or less.
\end{theorem}

\subsection{Error correction beyond {\boldmath{$t$}}}
For additive quantum codes, the syndrome measurement locates all
error equivalence classes, not only those with ``coset leaders'' of
weight $s<d/2$.  The same could be achieved with a series of clustered
measurements, by first going over all clusters of weight $s=t$, then
$s=t+1$, etc.  This ensures that the first located error has the
smallest weight.  In contrast, such a procedure will likely fail for a
non-additive code where the corrupted spaces $E_1(Q)$ and $E_2(Q)$ can
partially overlap if either $E_1$ or $E_2$ is non-correctable.  For
instance, the measurement in Example 6 may destroy the coherent
superposition if the actual error (e.g., $Z_i$, $i\neq2$) was not on
the second qubit.

Therefore, if no error was detected after $\binom{n}{t}-1$
measurements, we can continue searching for the higher-weight errors
only after testing the remaining size-$t$ index set.
 With a non-additive CWS
code, generally we have to do a separate measurement for each
additional correctable error of weight $s>t$.

\section{Conclusions}
\label{sec:conclusions}

For generic CWS and USt codes, we constructed a {\em structured
  recovery algorithm\/} which uses a single non-Pauli measurement to
check for groups of errors located on clusters of $t$ qubits.
Unfortunately, for a generic CWS code with large $K$
and large distance, both the number of measurements and the
corresponding complexity are exponentially large, in spite of the
exponential acceleration already achieved by the combined measurement.

To be deployed, error-correction must be complemented with some
fault-tolerant scheme for elementary gates.  It is an important open
question whether a fault-tolerant version of our measurement circuits
can be constructed for non-additive CWS codes.  It is clear, however,
that such a procedure would {\em not\/} help for any CWS code that
needs an exponential number of gates for recovery.  Therefore, the
most important question is whether this design can be simplified
further.

We first note that the group-based recovery [see
Theorem~\ref{theorem:error-detect}] is likely as efficient as it can possibly
be, illustrated by the example of additive codes in Sec.~\ref{sec:additive}
where this procedure is shown to be equivalent to syndrome-based recovery.
Also, while it is possible that for fixed $K$ the complexity estimate of
Theorem~\ref{theorem:complexity1} can be reduced in terms of $n$ (e.g., by
reusing ancillas with measured stabilizer values), we think that for a generic
code the complexity is linear in $K$.

However, specific families of CWS codes might be represented as unions
of just a few stabilizer codes which might be mutually equivalent as
in Eq.~(\ref{eq:ust-defined}), or non-equivalent
 \cite{Grassl-1997}.  The corresponding measurement complexity for
error detection would then be dramatically reduced.  Examples are
given by the quantum codes derived from the classical non-linear
Goethals and Preparata codes
 \cite{Grassl-Roetteler-2008A,Grassl-Roetteler-2008B}.
 
 Another possibility is that for particular codes, larger sets of
 correctable errors may form groups.  Indeed, we saw that for an
 additive code $((n,2^k,d))$, all error degeneracy classes form a
 large group of size $2^{n-k}$ which may include some errors of weight
 well beyond $t$.  Such a group also exists for a CWS code which is a
 subcode of an additive code.  There could be interesting families of
 non-additive CWS codes which admit groups of correctable errors of
 size beyond $2^{2t}$.  For such a code, the number of measurements
 required for recovery could be additionally reduced.

\section*{Acknowledgment}

This research was supported in part by the NSF grant
No.\ 0622242. Centre for Quantum Technologies is a Research Centre of
Excellence funded by Ministry of Education and National Research
Foundation of Singapore.  The authors are grateful to Bei Zeng for the
detailed explanation of the CWS graph construction.

\appendix
\section{Orthogonality of corrupted spaces}
\label{app:orthogonality}
As discussed in Sec.~\ref{sec:general}, for a general non-additive
quantum code ${\cal Q}$ and two linearly independent correctable
errors, the corrupted spaces $E_1({\cal Q})$ and $E_2({\cal Q})$ may
be neither identical nor orthogonal \cite{Calderbank-1997}.  However,
for CWS and USt codes it is almost self-evident that when
$E_1(\mathcal{Q})$ and $E_2(\mathcal{Q})$ do not coinside, they are
mutually orthogonal.  This
orthogonality is
inherited from the originating stabilizer code ${\cal Q}_0$.  In
particular, in some previous publications (e.g.,
Ref.~\cite{Cross-CWS-2009}) orthogonality is implied in the discussion
of degenerate errors for CWS codes.  However, to our knowledge, it was
never explicitly discussed for CWS or USt codes.  Since our recovery
algorithms for CWS and USt codes rely heavily on this orthogonality,
we give here  an explicit proof.

First, consider a stabilizer code ${\cal Q}_0$.  For any Pauli
operator $E\equiv E_1^\dagger E_2$, there are three possibilities:
({\bf i}) $E$ is proportional to a member of the stabilizer group,
$E\equiv \gamma S$, where $S\in\mathscr{S}_0$ and $\gamma=i^m$,
$m=0,\ldots, 3$, ({\bf ii}) $E$ is in the code normalizer
$\mathscr{N}_0$ but is linearly independent of any member of the
stabilizer group, and ({\bf iii}) $E$ is outside of the normalizer,
$E\not\in\mathscr{N}_0$.

Case ({\bf i}) implies that the space $E({\cal Q}_0)$ is identical to
the code ${\cal Q}_0$; the errors $E_1$ and $E_2$ are mutually
degenerate.  Indeed, for any basis vector $\ket i$, the action of the
error $E\ket i=\gamma S\ket i=\gamma\ket i$ just introduces a common
phase $\gamma$; any vector $\ket\psi\in {\cal Q}_0$ is mapped to
$\gamma\ket\psi$ and hence no recovery is needed.

In case ({\bf ii}) the operator $E$ also maps ${\cal Q}_0$ to itself,
but no longer identically.  Therefore, at least one of the two errors
$E_1$, $E_2$ is not correctable.  Indeed, in this case we can
decompose $E$ (see Sec.~\ref{sec:stabilizer-codes}) as the product of
an element $S\in\mathscr{S}_0$ in the stabilizer and logical
operators, i.e., $E\equiv i^m S \overline X^{\mathbf{a}}\overline
Z^{\mathbf{b}}$, where $m=0,\ldots, 3$ determines the overall phase.
While $S\in\mathscr{S}_0$ acts trivially on the code, the logical
operator specified by the binary-vectors $\mathbf{a}$, $\mathbf{b}$ is
non-trivial, $\wgt(\mathbf{a})+\wgt(\mathbf{b})\neq0$.  Using the
explicit basis~(\ref{eq:stabilizer-code-basis}), it is easy to check
that the error-correction condition (\ref{SNcondtion-correcting}) is
\emph{not} satisfied for the operators $E_1$, $E_2$.

Finally, in case ({\bf iii}) the spaces $E({\cal Q}_0)$ and ${\cal
  Q}_0$ are mutually orthogonal.  Indeed, since $E$ is outside of the code
normalizer $\mathscr{N}_0$, there is an element of the stabilizer
group $S\in\mathscr{S}_0$ that does not commute with $E$.  Therefore,
for any two states in the code, $\ket \varphi, \ket\psi\in{\cal Q}_0$,
we can write
\begin{equation}
\bra \varphi E\ket \psi=\bra \varphi E S\ket \psi=-\bra \varphi SE\ket \psi
=-\bra \varphi E\ket \psi, \label{eq:stabilizer-orthogonality}
\end{equation}
which gives $\bra \varphi E\ket \psi=0$, and the spaces ${\cal
  Q}_0$ and $E({\cal Q}_0)$ [also, $E_1(\mathcal{Q}_0)$ and
  $E_2(\mathcal{Q}_0)$] are mutually orthogonal.

Now, consider the same three cases for a USt
code~(\ref{eq:ust-defined}) derived from $\mathcal{Q}_0$.  In case
({\bf i}) the code is mapped to itself, $E({\cal Q})={\cal Q}$.  The
operator $E$ acts trivially on the code (and the errors $E_1$, $E_2$
are mutually degenerate) if $E$ either
commutes~(\ref{eq:translations-commute}) or
anti-commutes~(\ref{eq:translations-anti-commute}) with the entire set
of translations generating the code:
\begin{eqnarray}
 & &  \left(Et_j=t_jE, \; j=1,\ldots,K\right)
 \label{eq:translations-commute}\\
 & \text{or}&  \left(Et_j=-t_jE, \;  j=1,\ldots,K\right).
\label{eq:translations-anti-commute}
\end{eqnarray}
If neither of these conditions is satisfied, the error-correction
condition (\ref{SNcondtion-correcting}) is violated.  This is easily
checked using the basis $\ket{j,i}\equiv t_j\ket i$.

Similarly, in case ({\bf ii}), the code is mapped to itself, $E({\cal
  Q})={\cal Q}$, but the error-correction condition
(\ref{SNcondtion-correcting}) cannot be satisfied.

Finally, in case ({\bf iii}), the space $E({\cal Q})$ is either
orthogonal to ${\cal Q}$, or the error correction condition is not
satisfied.  The latter is true if $E$ is proportional to an element in
one of the cosets $t_j^\dagger t_{j'}\mathscr{S}_0$, where $j\neq j'$,
$1\le j,j'\le K$.  Then the inner product $\bra{j, i} E
\ket{j',i}\neq0$, $i=1,\ldots, k$, which contradicts the
error-correction condition~(\ref{SNcondtion-correcting}).  In the
other case, namely, when $E$ is linearly independent of any operator
of the form $t_j^\dagger t_{j'}S$, $j,j'=1,\ldots, K$,
$S\in\mathscr{S}_0$, $E$ must be a member of a different coset
$t_\alpha\mathscr{S}_0$ of the stabilizer $\mathscr{S}_0$ of the code
$\mathcal{Q}_0$ in $ \mathscr{P}_{n}$.  This implies orthogonality:%
$$
\bra{j,i}E\ket{j',i'}\equiv
\bra{i}t_j^\dagger\, E \,t_{j'}\ket{i'}=i^m
\bra{i}t_j^\dagger t_{j'}\, t_\alpha \ket{i'}=0,
$$
where $m=0,\ldots,3$ accounts for a possible phase factor.

Overall, as long as the error correction condition
(\ref{SNcondtion-correcting}) is valid for a USt code $\mathcal{Q}$
and the Pauli operators $E_1$, $E_2$, the spaces $E_1(\mathcal{Q})$
and $E_2(\mathcal{Q})$ either coincide, or are orthogonal.  Since
CWS codes can be regarded as USt codes originating from a
one-dimensional stabilizer code $\mathcal{Q}_0$,
[Sec.~\ref{sec:cws-defined}], the same is also true for any CWS code.


\begin{thebibliography}{38}
\bibitem{endnote38} Note that errors may act trivially on the code,
  i.e., they have no effect and cannot be detected. Nonetheless, we
  follow the majority of the literature and use the term ``detectable
  errors.''
\bibitem{endnote1} Note that the representatives of different error
  classes forming a group are not necessarily the minimum-weight
  representatives.
\end{thebibliography}
\end{document}